\documentclass[prl,aps,twocolumn,showpacs,floatfix]{revtex4}
\usepackage{graphicx,epstopdf}
\usepackage{bm}
\usepackage{amsmath}
\usepackage{amssymb}
\usepackage{amsfonts}
\usepackage{dcolumn}%
\usepackage{bm}
\usepackage{color}
\usepackage {ulem}
\usepackage{textcomp}

\def\mum{\nobreak\mbox{$\;$\textnormal{\textmu m}}}
\def\nls{\nobreak\mbox{$\;$\textnormal{nL/sec}}}

\newcommand{\be}{\begin{equation}}
\newcommand{\ee}{\end{equation}}
\newcommand{\bea}{\begin{eqnarray}}
\newcommand{\eea}{\end{eqnarray}}

\renewcommand{\deg}{{}^{\circ}}
\begin{document}
\title{Non-Newtonian viscosity of E-coli suspensions}
\author{J\'er\'emie Gachelin, Gast\'on Mi\~no, H\'el\`ene Berthet,  Anke Lindner, Annie Rousselet and \'Eric Cl\'ement}
\affiliation{
PMMH-ESPCI, UMR 7636 CNRS-ESPCI-Universities Pierre et Marie Curie and Denis Diderot, 10 rue Vauquelin, 75005 Paris,
France.}
\date{\today}
\begin{abstract}
The  viscosity of an active suspension of E-Coli bacteria is determined experimentally in the dilute and semi dilute regime using a Y shaped micro-fluidic channel. From the position of the interface between the pure suspending fluid and the suspension, we identify  rheo-thickening and rheo-thinning regimes as well as situations at low shear rate where the viscosity of the bacteria suspension can be lower than the viscosity of the suspending  fluid. In addition, bacteria concentration and velocity profiles in the bulk are directly measured in the micro-channel.
\end{abstract}
\maketitle

The fluid mechanics of microscopic swimmers in suspension have been widely studied in recent years. Bacteria \cite{Wu2000, Dombrowski2004}, algae \cite{leptos2010,Rafai2010} or artificial swimmers \cite{Paxton2004} dispersed in a fluid display properties that differ strongly from those of passive suspensions \cite{Baskarana2009Koch2011}. The physical relationships governing momentum and energy transfer as well as constitutive equations vary drastically for these suspensions \cite{Hatwalne2004, Chen2007}. Unique physical phenomena caused by the activity of swimmers were recently identified such as enhanced Brownian diffusivity \cite{Wu2000, Chen2007, Mino2011, Underhill2008} uncommon viscosity \cite{Sokolov2009, DiLeonardo2010,Rafai2010}, active transport and mixing \cite{Kim2007} or  the extraction of work from isothermal fluctuations \cite{Sokolov2010, DiLeonardo2010}.  The presence of living and cooperative species may also induce collective motion and organization  at the mesoscopic or macroscopic level \cite{Gregoire2001,Shelley 2012} impacting the constitutive relationships in the semi-diluted or dense regimes.\\
\begin{figure}[htb]
\begin{center}
\includegraphics[width=.5\textwidth]{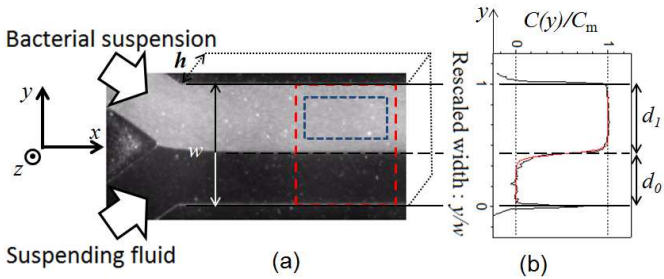}
\caption{ Experimental set-up. (a) Time averaged image of the micro-channel ($W=600\mum$) for $Q=10\nls$ and volume fraction $\phi=0.35\%$. The red and blue frames indicate the measurement areas. (b) Concentration profile $C(y)$ normalized by the maximal concentration $C_M$ (black line) and error function fit used to determine the interface position (red line).}
\label{Figure1}
\end{center}
\end{figure}

The E.Coli bacterium possesses a quite sophisticated propulsion apparatus consisting of a collection of  flagella (7-10  $\mu m$ length) organized in a bundle and rotating counter-clockwise \cite{Berg2004}. In a fluid at rest, the wild-type strain used here has the ability to change direction by unwinding  some flagella and moving them in order to alter its swimming direction (a tumble) approximately once every second \cite{Darnton2007}. In spite of the inherent complexity of the propulsion features, low Reynolds number hydrodynamics impose a long range flow field which can be modeled as an effective force dipole. Due to the thrust coming from the rear, E.coli are described as "pushers", hence defining a sign for the force dipole which has a crucial importance on the rheology of active suspensions \cite{Hatwalne2004}. For a dilute suspension of force dipoles, Haines et al \cite{Haines2009} and Saintillan \cite{Saintillan2010} derived an explicit relation relating viscosity and shear rate. They obtained  an effective viscosity similar in form to the  classical Einstein relation for dilute suspensions :  $\eta = \eta_0 (1+K \phi)$ ($\eta_0$ being the suspending fluid viscosity and $\phi$ the volume fraction). These theories predict a negative value for the coefficient $K$ for pushers at low shear rates, meaning the suspension can exhibit a lower viscosity than the suspending fluid. The theoretical assessment of  shear viscosity relies on an assumed statistical representation of the orientations of the bacteria, captured by a Fokker-Plank equation and a kinematic model for the swimming trajectories \cite{Jeffrey1922Bretherton1962, Zoettl2012}.\\

%
%
Despite the large number of theoretical studies, few experiments have been conducted. With \textit{bacillus subtilis} (pushers) trapped in a soap film Sokolov et al.\cite{Sokolov2009} have shown that a vorticity decay rate could be associated with a strong decrease of shear viscosity in the presence of bacteria. For algea (pullers), Rafai et al. \cite{Rafai2010} have shown that the effective viscosity measured in a classical rheometer is  larger than the viscosity of the corresponding dead (passive) suspension. However, no measurements of the viscosity of a dilute suspension of pushers under controlled shear conditions exist to date. This is mainly due to the fact that one has to assess small viscosities near the viscosity of water at very small shear rates to probe the theoretical predictions. These parameters are typically outside of the resolution of a classical rotational rheometer and have thus made these measurements inaccessible. In this letter, we present the first measurements of the shear viscosity of a suspension of pushers using a microfluidic device to overcome these difficulties and obtain the relative viscosity of an active suspension  for a large range of shear rates and bacteria concentrations. Our set-up also allows direct visualization of the flow as well as the spatial distributions of bacteria in the flow. \\
The wild type \textit{E. coli} W used here are prepared following the experimental procedure described in Ref. \cite{Berke2008, Mino2011}. The strain is grown overnight in rich medium (LB). After washing, it is transferred into MMA, a motility medium  supplemented with K-acetate (0.34 mM) and polyvinyl pyrolidone (PVP: 0.005\%). The sample is then incubated for at least one hour. To avoid bacteria sedimentation, Percoll is mixed with MMAP 1vol/1vol (isodense conditions). We verified that under these conditions, the suspending fluid is Newtonian (viscosity $\eta=1.28\times 10^{-3}$ Pa\,s at  22$^{\circ}$C). All experiments are performed at a fixed temperature $T=25\deg C$.\\
To obtain the shear viscosity, we adapted a microfluidic device \cite{Guillot2006} to compare the Newtonian viscosities of two liquids. The device is  a Y-shape Hele-Shaw cell of height $h$, such that each branch receives a different fluid (resp fluid $0$ of viscosity $\eta_0$ and fluid $1$ of viscosity $\eta_1$ (see fig.\ref{Figure1}).  Both flows are driven at an identical flow rate $Q$. In a Hele-Shaw approximation, the velocity profile in the $z$ direction can be described by a parabolic flow profile and the dominant shear rate occurs in the $z$ direction. Under these conditions and for a viscosity ratio near one, the position of the interface between both fluids at steady state i.e. $d_1/d_0$ is determined by the viscosity ratio $\eta_1/\eta_0$ \cite{Guillot2006}. Guillot {\it et al.} \cite{Guillot2006} have also used this approach for non-Newtonian fluids, where it corresponds to the measurement of an apparent viscosity. Here we will follow the same approach and will verify our method subsequanty. The suspension of bacteria is flown into one arm  and the suspending Newtonian fluid into the other arm. The interface position is then measured at various flow rates $Q$. The experimental data is presented as a function of the maximum shear rate obtained by assuming a parabolic flow profile $\overset{.}{\gamma}_{M}=(6 Q)/(h^{2}d_1)$ where  $d_1$ is the lateral width occupied by the suspension.  The relative viscosity $\eta_r=$  is then  :
\begin{equation}
\eta_r=\frac{\eta _{1}}{\eta _{0}}=\frac{d_{1}}{d_{0}}
\label{equation:visc}
 \end{equation}
 This micro-fluidic device has the advantage of measuring a viscosity ratio and provides very good resolution of the suspension viscosity independent of its absolute value or the applied shear rate. Special care has to be taken that both branches are fed at the same constant flow rate $Q$. We use a very high precision two-syringe pump from \textit{nemeSYS} and a precision syringe (Hamilton Gastight 1805RN) of a very small volume (50$\mu L$) allowing to impose identical and very small flow rates (down to $Q=0.5~\nls$) on both arms. Finally, the position of the sample region (indicated by the red rectangle on figure \ref{Figure1}) is chosen in such a way as to be in steady state conditions (reached at a distance of approximately $600~\mu m$ from the junction of the inlet channels) while avoiding significant mixing. \\
The Y-shape channel was fabricated completely from PDMS using a soft-lithography technique. The channel thickness is $h=100~\mu m$. The main channel width is $w=600~\mu m$ and the two inlet branches widths are $w/2$. Inlets are connected by $500~\mu m$ diameter tubes to the two-syringe pump. The total length of the main channel is $40 mm$. Suspensions were prepared (see \cite{Mino2011}) with a number of bacteria per unit volume $n$ in the range $1.9~10^{12} L^{-1}<n<25.6~10^{12} L^{-1}$. The volume fraction is estimated using the space occupied by the body of each bacteria $v_b=1~\mu m^3$ such that $\phi=n v_b$, yielding a range of $0.19<\phi<2.56\%$. Note that other definitions of the volume fraction, for example based on the length of the body or including the length of the flagella would lead to significantly higher volume fractions. The flowing suspension was observed using an inverted microscope (Zeiss-Observer, Z1) connected to a digital camera PixeLINK PL-A741-E ($1280~X~1024 pix^{2}$) capturing videos at a frame rate of 22 images/s. Low magnification (2.5x) allowed  an extended view of the channel (see Fig.\ref{Figure1}).\\
%
%
%
\begin{figure}[htb]
\begin{center}
\includegraphics[width=.45\textwidth]{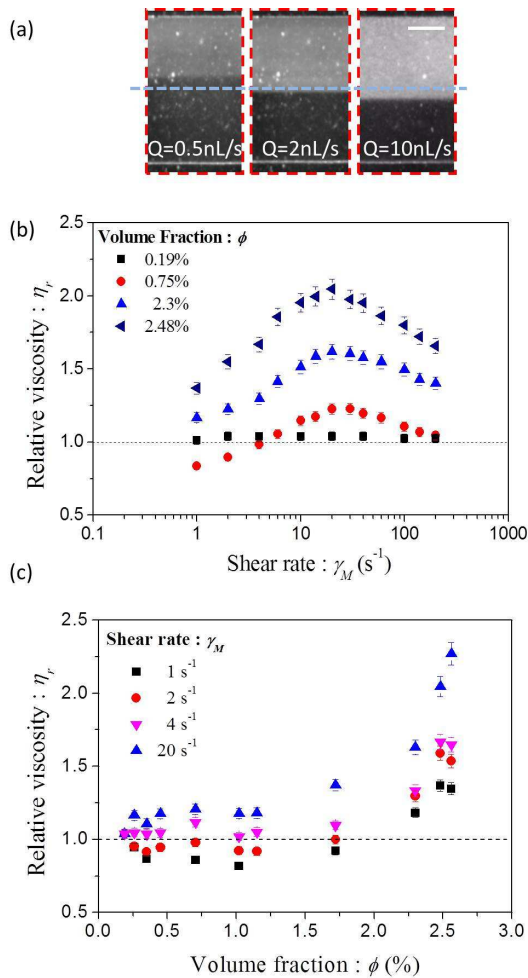}
\caption{Rheology curves. Fig (a) are 3 averaged pictures of the bilaminar flow for flow rates of $Q=0.5~nL/s$, $Q=2~nL/s$ and $Q=10~nL/s$ respectively for a $\phi=0.35\%$ bacteria suspension. These images are averaged over 120 images at 22im/s and observed with a 2.5x magnification using phase contrast. The dotted line represents the mid-position in the channel width. The scale bar corresponds to $200\mum$. (b) Relative viscosity versus mean shear rate at several volume fractions (c) Relative viscosity versus volume fraction at several shear rates. The errors bars are estimated using the detection error of the interface position.}
\label{Figure2}
\end{center}
\end{figure}
%
During an experiment we begin with an initial flow rate of $Q=0.5~1\nls$, and then the input flow is  increased  step by step up to $Q=100~\nls$. Note that we have verified for all experiments that identical results are obtained when subsequently decreasing the flow rate. In fig \ref{Figure1}, the shape of the interface obtained by averaging over $120$ successive images is displayed for the measurement area. To quantitatively determine the interface position, we measure the mean bacteria density across the channel width $\overset{-}{n}(z) $  using the logarithm of the profile of the average intensity $C(y)=\ln(\langle I\rangle(y)/\langle I\rangle_0$ as a measure of the concentration, with $\langle I\rangle_0$ is the mean intensity in the absence of bacteria. The average is calculated in the x-direction over a distance of $600~\mu m$ (see red rectangle in fig.\ref{Figure1}). The concentration $C(y)$ is fitted with an error function $\mbox{erf}(y)$ to obtain the interface position $y_I$.  Once the interface position is determined, we extract the relative viscosity according to equation (\ref{equation:visc}) and associate it with $\overset{.}{\gamma} _M$.  \\
Experimental observations are presented in fig.~\ref{Figure2}(a) displaying averaged images for flow rates of $Q=0.5~\nls$, $Q=2~\nls$ and $Q=10~\nls$ for a suspension at a volume fraction of $\phi=0.35 \%$. When increasing the flow rate the interface position changes from a value above the mid-position to a value below the mid-position indicating a change in the suspension viscosity from greater then the viscosity of the suspending fluid to smaller. Quantitative measurements are given in fig.\ref{Figure2}(e) showing the relative viscosity $\eta_r$ of the suspension as a function of the maximum shear rate for various concentrations $\phi=0.19,~0.75,~2.3$ and $2.48~\%$. The error bars are estimated using the uncertainty in determining the interface position and in marked contrast to classical rheology, the error does not increase for lower shear rates, once again demonstrating the strength of our micro-rheometer. For the lowest concentration the viscosity of the Newtonian suspending fluid is recovered for all shear rates validating our rheological device. For all other concentrations, the curves display the same qualitative behavior. At low shear rates, we observe shear thickening  and at higher shear rates, shear thinning occurs.  For all of these cases, the shear thinning and shear thickening character of the active suspensions are weak and power law indices $n$ (using $\eta=K\overset{.}{\gamma}^{(n-1)}$) close to one are found for both regimes for all concentrations.  At all concentrations tested, a maximum viscosity is observed at a value around $\overset{.}{\gamma} _M=20 s^{-1}$.  The non-monotonic behavior as a function of shear-rate is in agreement with the results of Saintillan \cite{Saintillan2010}  obtained for slender bacteria. Note that the low shear rate plateau predicted by the theory could not be observed in our experiments due to practical limitations of our device (minimum flow rates $0.5 \nls$).  At low volume fractions and low shear rates, we observe a decrease of $\eta_r$ below one which agrees with the theoretical predictions of Haines et al. \cite{Haines2009} and Saintillan \cite{Saintillan2010} of this striking effect. Figure \ref{Figure2}(c) shows the relative viscosity as a function of the volume fraction $\phi$ at various shear rates $\overset{.}{\gamma _M}=1,~2,~4$ and $20~s^{-1}$. A decrease in viscosity is observed for the small shear rates, followed by a sharp increase of viscosity for all shear rates, corresponding to a semi-dilute regime. In our case this regime is observed for concentrations above approximately $1\%$. Similar behavior was also observed by Sokolov et al.\cite{Sokolov2009} using vortex decay in a suspension of Bacilus subtilis confined in a soap film and has been predicted by Ryan et al. \cite{Ryan2011} in their simulations. \\
So far we have measured the shear viscosity of the suspensions as a function of the maximum shear rate assuming a parabolic velocity profile. One of the advantages of our microfluidic device is that we can access the local velocity and bacteria concentration profiles as a function of the channel height directly. Bacteria moving in the flow were visualized with a high magnification objective (40x, phase contrast) allowing the position of the bacteria to be monitored at various heights $z$ (field depth $3~\mu m$). Videos were taken using a highspeed camera (Photron FastCam SA3, resolution 1024x1024pixels, shutter speed ($1/500 s$), frame rate $1/50 s$ for $Q=0.5, 1 nL/s$ and $1/500 s$ for $Q=10 nL/s$). To monitor the flow velocity,  we suspended  a very low concentration of $2 \mu m $ density-matched latex beads as passive tracers (density $1.03 g/cm^3$).
Here we present selected results for the volume fraction $\phi=0.75~\%$ at which the reduction of the relative viscosity below $1$ is observed at three flow rates ($Q=0.5, 1, 10~\nls$ corresponding respectively to maximum shear rates $\overset{.}{ \gamma_{M}}=1,2,20~s^{-1}$. Videos were taken  in a region situated in the middle between the interface and the side wall (see blue frame of fig. \ref{Figure1}). To reduce the detection noise, a bacterium is retained only if it is detected on at least two consecutive frames. The flow velocities were computed for particles moving in the focal plane. In Fig. \ref{Figure3}a) we see that the concentration profiles are similar in shape to those published previously by \cite{Berke2008} in the absence of  flow, i.e. a quasi-constant density and a strong density increase within $10 \mu m$ of the wall, hence probing a trapping effect that persists over the range of shear rates explored. The calculated mean concentration is represented by the dashed line in fig. \ref{Figure3}a and is in agreement with the measured bulk concentration. It is important to note that the concentration profiles are identical for the various shear rates. The change in viscosity observed is therefore not due to a change in concentration within the microchannel. The velocity profiles $V_{x}(z)$ are displayed in Fig.(\ref{Figure3}b). Within the precision of our experimental setup, a deviation from a parabolic velocity profile is not observed justifying the approximation used for our analysis. Note also that this finding is in agreement with the weak shear thinning or shear thickening character of our suspensions, that predict only small changes in the velocity profile, not detectable in our experiments. For each flow rate the curves presented in fig. \ref{Figure3}b) agree within $3 \%$ with the theoretical predictions for a rectangular cell of the aspect ratio of (1/6) \cite{White2005}.\\
%
%
%
\begin{figure}[htb]
\begin{center}
\includegraphics[width=.4\textwidth]{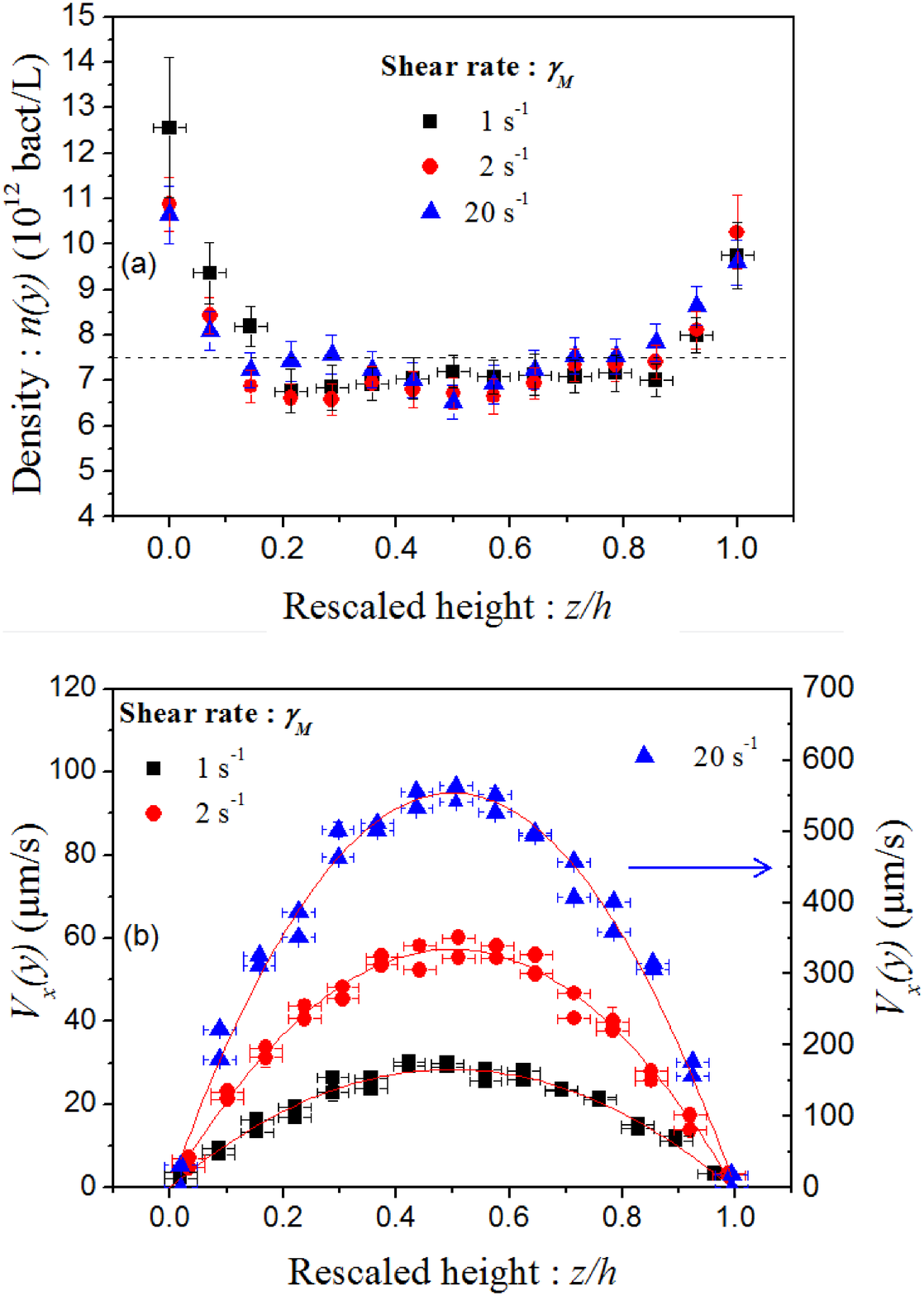}
\caption{Bulk profile measurements for a volume fraction $\phi=0.75~\%$  and shear rates $ \overset{.}{ \gamma_{M}}=1,2, 20~s^{-1}$. (a) Bacteria density profile $n(z)$. The horizontal dashed line represents the mean density: $ \overset{-}{n}=7.5 \mbox{bact}/L$. (b) Velocity profile of passive tracers $V_{x}(z)$.}
\label{Figure3}
\end{center}
\end{figure}

In conclusion, using a micro-fluidic device, we have measured for the first time the effective shear viscosity of a suspension of  "pushers" ( E-coli bacteria) over a large range of  shear rates($ 1-200~s^{-1}$) under controlled shear conditions in the dilute and semi-dilute regime. We confirmed an important prediction for the rheology of pushers:  the active viscosity can be lower than the viscosity  of the suspending fluid at low shear rates \cite{Hatwalne2004,Haines2009,Saintillan2010}. In the dilute as well as in the semi-dilute regime, we observed a shear-thickening behavior at lower shear rates followed by a shear-thinning regime at higher shear rates. The  viscosity  maximum  is observed at a shear rate on the order of the inverse of the time a bacterium needs to swim over a distance of its own size and this value seems independent of the bacteria concentration.  These results are consistent with  the theoretical calculation of Saintillan \cite{Saintillan2010}.
Direct measurements of the bacteria concentrations and  flow velocities confirm that the bacteria distribution remains homogeneous and the parabolic  flow profile is not significantly modified  by the weak non-Newtonian character of the active fluid. This last result validates the use of the Newtonian approximations for the extraction of our viscosity measurements. In the semi-dilute regime (here for volume fractions greater than $ 1 \%$), we observe a strong increase of the viscosity consistent with  numerical simulations by Ryan et al.\cite{Ryan2011}. Our results represent the first experimental validation of the non-Newtonian rheology of an active suspension of pushers under controlled shear conditions. \\

\textit{Acknowledgment.} We acknowledge the financial support of the Pierre-Gilles de Gennes  Foundation and the SESAME Ile-de-France research grant :" Milieux actifs et stimulables". We thank Profs R.Soto, D.Saintillan, I.Aronson and M. Alves for enlightening scientific discussions and C. Davis for a careful reading of the manuscript.\\
%

\end{document}